# Characterizing the Immaterial: Noninvasive Imaging and Analysis of Stephen Benton's Hologram *Engine no. 9*

Marc Walton[1], Pengxiao Hao[1], Marc Vermeulen[1], Florian Willomitzer[2], and Oliver Cossairt[2]

1. Center for Scientific Studies in the Arts, Northwestern University, 2145 Sheridan Road Evanston, Il, USA
2. Department of Computer Science, Northwestern University, 2145 Sheridan Road Evanston, Il, USA

**Abstract**
Invented in 1962, holography is a unique merging of art and technology. It persisted at the scientific cutting edge through the 1990s, when digital imaging emerged and supplanted film. Today, holography is experiencing new interest as analog holograms enter major museum collections as bona fide works of art. In this essay, we articulate our initial steps at Northwestern's Center for Scientific Studies in the Arts to describe the technological challenges on the conservation of holograms, emphasizing their nature as an active material—a holographic image requires user interaction to be viewed, and the materials are delicate and prone to deterioration. Specifically, we outline our methods for creating digital preservation copies of holographic artworks by documenting the wavefront of propagating light. In so doing, we demonstrate why it remains challenging to faithfully capture their high spatial resolution, the full parallax, and deep depths of field without terabytes of data. In addition, we use noninvasive analytical techniques such as spectral imaging, X-ray fluorescence, and optical coherence tomography, to provide insights on holograms' material properties. Through these studies we hope to address current concerns about the long-term preservation of holograms while translating this artform into a digital format to entice new audiences.

**Introduction**

> The fragment breaks loose with a snap, sending Luke tumbling head over heels. He sits up and sees a twelve-inch three-dimensional hologram of Leia Organa, the Rebel senator, being projected from the face of little Artoo. The image is a rainbow of colors as it flickers and jiggles in the dimly lit garage. Luke's mouth hangs open in awe.[1]

This scene of the Princess Leia hologram from George Lucas's film *Star Wars* perfectly embodies the futuristic zeitgeist of the mid-1970s. Indeed, anyone who has seen a large-format analog hologram in person is familiar with the beguiling realism it offers. In the early days when the technology was still new and fresh, the impression of a complete three-dimensional (3D) picture frozen in time could be unnerving, as illustrated by the apocryphal story of Cartier's 1972 Fifth Avenue display of a hologram depicting a hand holding jewelry (fig. 1) causing an elderly woman to attack the virtual arm floating in thin air.[2] Since this heyday of holography, many of these works have been collected by museums. Notably, the MIT Museum holds one of the largest archives with over two thousand holograms originating from the now defunct Museum of Holography (1976–92).[3] More recently, in 2018 the Art Institute of Chicago obtained *Huddle* by Simone Forti, a two-hundred-degree multiplex hologram.[4] In 2019, the J. Paul Getty Museum acquired

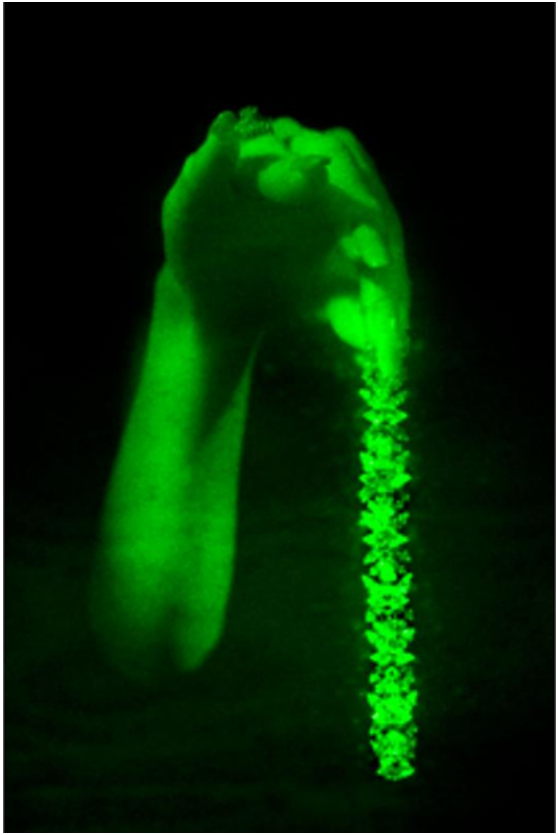

*Figure 1. Robert Schinella, Hand and Jewels, 1980s–90s. Holograms; 12 × 16 in. to 18 × 24 in. (30.48 × 40.64 cm to 45.72 × 60.96 cm). MIT Museum collection. Photo adapted from "[Cartier] Hand and Jewels."*

105 hologram pieces related to the *C-Project*, a collection of holograms produced in the 1990s by major contemporary artists to promote holography as an artistic medium.[5] Yet, while scientists, engineers, and artists have been creating holograms for over sixty years, very little work has been done to document the wide variety of imagery stored within them, nor to study how to preserve the large body of holograms housed at museums and galleries.

Holography, though closely related to its cousin photography, is a vastly more complicated medium, with image information recorded in a different way than conventional photographic processes. Conventional photography relies on the projection of light rays incident onto a photosensitive surface. This optical arrangement results in a direct, and easily understandable, relationship between the density of the silver grains and the observed contrast of a photographic image at the macroscopic level. A hologram, on the other hand, relies on recording the interference produced by the wavefront reflected from an object or scene. Also, because holograms rely on the wave nature of light, they can only be recorded using coherent light sources, such as a laser (light amplification by stimulated emission of radiation). The 3D image of a hologram is therefore encoded in the emulsion as nanometer-spaced interference fringes, analogous to a diffraction grating. These fringes bear no physical resemblance to the wavefront image that is formed only once the correct angle and wavelength of light is reflected off its surface or transmitted through its thickness. Perhaps it is the ambiguous nature of where the image forms in the physical material that accounts for the lack of conservation efforts and scientific study of holographic materials. It is intimidating to even suggest where one should begin when considering the material aspects of an object formed from light. Also, what exactly is one attempting to preserve with something so immaterial? Whatever the reasons, knowledge about holographic materials and the technical processes of making them remains ambiguous, and conservation approaches lack clarity and precedence.[6]

**Holograms as Active Matter**
One of the primary problems of the material study of holograms is that a hologram itself is not matter at all. Put another way, a hologram is a 3D object entirely sculpted by light interference. The only physical material in a hologram is the photographic diffraction grating, and arguably this encoded structure is no more than a vehicle of delivery for the image content. Therefore, there is only an incidental relationship between what one observes in a hologram and the physical materials constituting a holographic artwork. This is a very different situation than one would find in most any other artistic medium wherein appearance is the direct result of light being absorbed, scattered, or reflected in or off the materials constituting the object itself. Rather, a hologram is akin to the effect of light in a gothic cathedral, where the material properties of colored glass are subservient to its primary function of filtering wavelengths of natural light, thus synthesizing an illusionistic environment which elicits a metaphysical response—what Abbot Suger referred to as the *lux nova* in the twelfth century.

So, on the face of it, the conservation of holograms might seem fairly straightforward. Since the carrier of the information (the photographic emulsion) is not as important as the image, one approach to conservation is to transfer image content ad infinitum onto new emulsions or displays as the entropy and decay of the original sets in. In fact, this approach is already in practice, with the most notable case being Louise Bourgeois's holograms at the Museum of Modern Art in New York, which were remade after the original film had deteriorated.[7] Here, the restoration efforts of the artist's images of architecture, body parts, and other assorted objects went so far as to transfer the images onto a more stable glass substrate! However, this "ship of Theseus" approach to the conservation of active matter, wherein the persistence of the physical materials is not a requirement, diminishes the agency of holographers who have greatly concerned themselves with controlling color, sharpness, and indeed, the overall visual effects of what they created through the manipulation of the photographic emulsion. Such concerns also extend to the illumination of holograms, in which artists chose a huge variety of approaches that were often optimized for the particular installation of a given artwork. For instance, the first time Simone Forti exhibited her *Angel* in 1976, she used a candle as the source.[8] While beyond the scope of this present essay, these concerns about how holograms were intended to be played back also needs better documentation, with an emphasis placed on the nature of the ambient light in the viewing area and the properties of the illuminating element.

In the remainder of this essay we will unpack what constitutes the material structure of a reflection hologram and, through a case study, explore how one holographer, Stephen Benton, crafted a hologram of a toy train which itself paid homage to the history of making holograms.

**Holograms as Physical Matter**
Invented in 1948 as a way to improve the resolution of electron microscopes, holography wasn't fully realized until the proliferation of the laser in the 1960s and 1970s. Shortly after scientists and engineers developed the first techniques to record analog holograms using lasers in a laboratory setting, the potential of this new medium captured the imagination of artists fascinated with the possibility of working with a new canvas that enabled them to sculpt directly with light. The initial technical draw of the hologram was the futuristic promise of a realistic 3D display. In many respects, such a 3D display (think of the fictitious Princess Leia hologram) couldn't be realized using just analog film—a fact understood by most of the original holographers—but the conceptual and mathematical basis for such a device was in place. Only now, nearly sixty years after the invention of the hologram, is the technology meeting this initial vision with developments like Google's Project Starline.[9] Artists, however, from the start of film-based holography, were fascinated with the new possibilities for abstract expression, 3D composition, color, texture, and motion. The first holographic artists emerged in the late 1960s, shortly after holographic photography was popularized by Emmett Leith and Juris Upatnieks, reaching the greatest numbers in the late 1970s and 1980s, then starting to decline after the popularization of the personal computer and digital graphical media.

The physical material of analog holography is high-resolution photosensitive film specifically designed to record interference fringes with a resolution close to the wavelength of light. When looking at a large-format hologram (e.g., one square meter), the information density is quite obvious: 3D images are displayed with stunning realism and, in some cases, almost indistinguishable from a real object.[10] Even today's massive twenty-to-forty-megapixel digital sensors lack the density to record the information embedded in an analog hologram, thus relegating practical digital holography to the dustbin, at least for now. Given this background, technical study of holograms may include everything from evaluating deterioration of the emulsion to reverse engineering the techniques of making, as well as displaying and documenting their dynamic visual effects. However, given the large variations in holographic recording setups, processing chemistries, and the sheer density of visual information contained in a hologram (fig. 2), what is the basic information required to document a hologram's state of preservation?

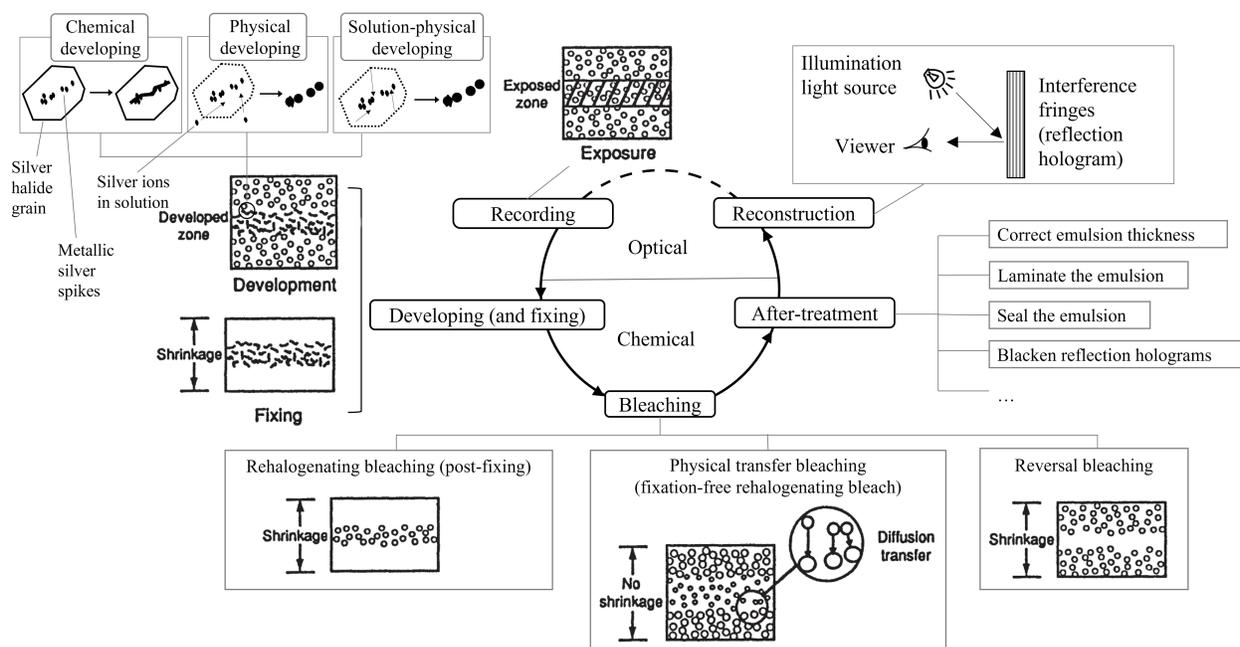

*Figure 2. Making and viewing holograms include optical and chemical steps: recording, developing (and fixing), bleaching, after-treatment, and reconstruction. Holographers may apply different variations of procedures in each step. Image adapted from Bjelkhagen, Silver-Halide Recording Materials.*

To address this question, we first must consider the key procedures in making and viewing holograms. One simple example of a recording setup is the so-called in-line "Denisyuk" reflection holography as shown in figure 3.[11] The laser beam emitted from the source passes through a beam expander to hit the holographic plate and the object. When this beam (reference beam) and the light scattered from the object (sample beam) interfere, the holographic plate records these interference fringes.

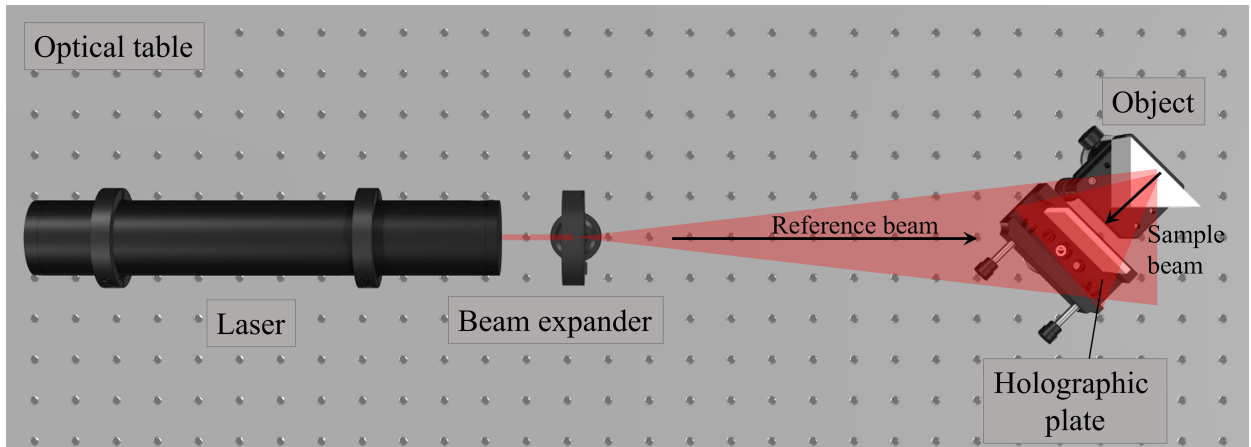

*Figure 3. Diagram of the optical setup of "Denisyuk" reflection holography. The diagram was drawn using the 3DOptix optical simulation and design software.*

The best display of a hologram, without distortion, requires an external light source to match the reference beam position as it is set in the recording setup. In practice, this beam position is found by trial and error to identify the "sweet spot" when the hologram is at its brightest. As such, reconstruction geometries may vary in each showing of a given hologram and, correspondingly, the reconstruction may change in appearance. Obvious concerns arise from these uncertainties in reconstruction, such as the extent to which the position of the light affects the visual experience of the hologram.

Perhaps the most notable characteristics of holograms is the realistic 3D space they occupy. Because the scene in a hologram moves with the viewing position, the viewer is often found dancing around it in an effort to take in its entirety. Benton once observed that "looking at this magical surface should be like looking through a window: we should see a three-dimensional image of the scene floating behind it with perfect realism, just as it would have looked if we saw the scene itself. We have created a 'window with a memory.'"[12] Obviously, conventional single-shot photodocumentation of a holographic image will fail to capture this "window of memory." Therefore, the documentation of a hologram calls for another approach that is compatible with this dynamic space, as will be discussed in greater detail below.

In addition to the optical appearance of holograms, the material properties relating to development and bleaching are important aspects to consider. Holographic processing chemistries are nearly identical to those used in conventional silver gelatin photography.[13] In fact, early holograms in the United States borrowed from thirty-five-millimeter Kodak films and processing.[14] However, these holograms were visually dissatisfying, possessing low brightness caused by the opaque silver particles and limited angular resolution (parallax) due to the relatively large grain sizes. To overcome these limitations, specialized holographic plates with smaller grains were produced, and developing agents were modified. Likewise, to increase the brightness of reflected or transmitted light, bleaching chemistries were innovated wherein silver halides (AgX, where X = $^-$I, $^-$Br, or $^-$Cl) were used in the finished emulsion rather than silver metal—a so-called phase grating, which produces a wavefront through refraction instead of scattering and absorption inherent to an amplitude grating. Because of these chemistries, conservation concerns for holograms go beyond conventional problems of photographic materials. They include complications associated with AgX particles that remain light-sensitive and will yellow with prolonged exposure to light as well as residual acidity associated with chemical components left over from the bleaching reagents.[15]

As shown schematically in figure 2, the following three types of developers have traditionally been applied to holography:[16]

> 1. *Chemical (direct) developers* are the most common developing agents which reduce silver halide in contact with exposed silver. The "developed" silver grains have "filament-like" shapes.
>
> 2. *Physical developers* function in two steps. Fixation first dissolves and removes all unexposed AgX from the emulsion. Then, exposed particles of silver act as seeds and nucleate fine clusters of metallic silver.
>
> 3. *Solution-physical developers* do not contain silver components. Instead, they dissolve the unexposed AgX grains to produce silver ions, which redeposit around the exposed silver to form fine colloidal clusters. More ideal for emulsions with finer grains, these developers were more common in the Soviet Union than in the West.

Bleaches generally fall into these three categories:[17]

> 1. *Rehalogenating (conventional) bleaches* are used together with a fixing agent. First, unexposed AgX is washed away. Then, developed silver grains are converted into a halide salt. Rehalogenating bleaches cause shrinkage of emulsion since unexposed AgX is lost in fixing.
>
> 2. *Physical transfer bleaches* are used without fixation. This type of bleach converts the metallic silver to AgX and migrates the silver ions to the unexposed grains. Physical transfer bleaches conserve the emulsion thickness because the least amount of silver components is lost.
>
> 3. *Reversal bleaches* are used without a fixer. With these bleaches, metallic silver is preferentially dissolved, leaving unexposed AgX in the final film. Because metallic silver is lost in bleaching, reversal bleaches also cause shrinkage of emulsion.

Although bleaching improves the brightness of the hologram, the process often causes shrinkage of the emulsion. Accordingly, the fringe spacings also decrease to produce a blue shift in the hologram relative to the wavelength of the laser illuminator. For instance, a reflection hologram recorded using a red He-Ne laser (633 nm) usually retains a green-yellow tint rather than the red of the laser. To compensate for (or indeed, sometimes induce) these color shifts, holographers traditionally tried to optimize such processing chemistries as will be discussed in more detail in the next section.

**Stephen Benton and His *Engine no. 9* Series**

Stephen Benton (1941–2003) was a pioneer in holography. Prolific in his invention of novel 3D displays and optics, Benton worked on holography over his entire career at Polaroid and then MIT.[18] Among his fifteen patents, his rainbow (white light transmission) holograms resulted in the ubiquitous embossed security holograms found to this day on credit cards and currency. Likewise, his dynamic holographic video system ushered in the idea of holographic TV. These achievements obscured many other less recognized, yet still key, findings. Also, though he would not have called himself an artist, he was an active promoter of the holographic field and remained a bridge between the engineering and art communities until his death.

In the United States, off-axis transmission holograms had dominated holographic research. Reflection holography, which was invented by Yuri Denisyuk in 1961 in the Soviet Union, was slow to take hold.[19] Yet compared with transmission holograms, reflection holograms held a significant advantage in that they required only a white light source for illumination instead of a laser. These different optical geometries produced shorter fringe spacings resulting in low brightness, inconsistent color shifts in hologram to hologram, and limited viewing angles due to the use of low-resolution emulsion. Such limitations prompted Benton to explore optical setups and processing chemistries to create a new type of reflection hologram with low noise, high brightness, and deep viewing depth, his *Engine no. 9* series.[20] As shown in figure 4, one example from this series depicts a toy train made in homage to the first hologram created by Leith and Upatnieks.

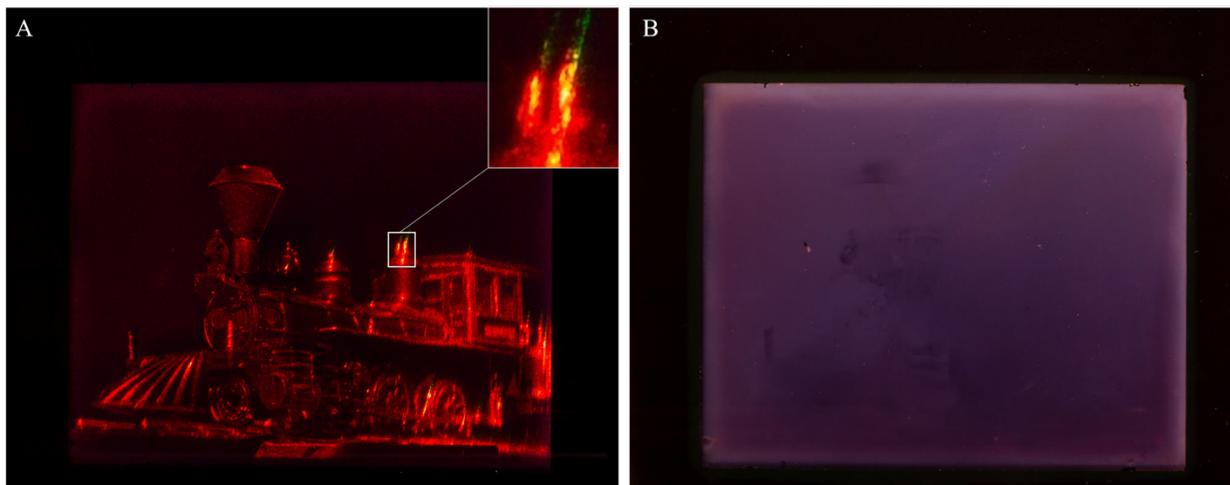

*Figure 4 Stephen Benton and Polaroid staffs, Engine no. 9 (labeled "Benton 12"), 1976–77. Hologram; 6.5 × 6.5 in. (16.51 × 16.51 cm). MIT Museum collection. A, with reconstruction. The insert, the whistle of the locomotive in magnification, exhibited colors of rainbow. B, without reconstruction.*

Considering the twelve copies of the hologram studied here, the red color and the appearance of the train (fig. 4A) exhibit considerable variation across the series. Also, evidence of deterioration is found on most of the holograms (fig. 4B), including loss of emulsion on edges and corners, dark patterns in regions of overexposure, and, to be discussed further, whitish "tide lines" along the edges. Although processing chemistries with a similar principle had been developed previously, particularly in the former Soviet Union, Benton pioneered the use of "IEDT (intra-emulsion diffusion-transfer) bleach" for this series.[21] Applying a combination of a chemical developer, a solution-physical developer, and a rehalogenating bleach in sequence resulted in minimal shrinkage to the emulsion during processing.[22] In addition, these steps generated a higher diffraction efficiency and thus brightness "by preserving the polarizable volume of the emulsion."[23] Therefore, by experimenting with IEDT, Benton was able to maintain the color of the reflection holograms and produced one of the brightest reflection holograms ever attained.

To explore the *Engine no. 9* series, we focused on two techniques to characterize the optical properties of the holograms: light-field captures to document the wavefront, or the 3D space occupied by the hologram, and spectral analysis to quantitatively characterize the color of reflection holograms. Additional characterization techniques such as optical coherence tomography (OCT) and X-ray fluorescence (XRF) will be briefly discussed in a later section.

To view the holograms comprised in the *Engine no. 9* series, we used a 3.5-mm-diameter white light LED emitter. Hans Bjelkhagen and David Brotherton-Ratcliffe suggested two general rules for replaying holograms. First, to avoid image blurring, they suggested the use of a small source size and long distance to the hologram to approximate a collimated beam. Second, to avoid aberrations, such as distortion of color and shape, the suggested requirement was to match the illumination geometry to the recording setup.[24] Since the exact recording geometry of the *Engine no. 9*s was unknown (Benton's laboratory records were destroyed), we fixed the position of the light source to allow for its maximum distance to the hologram, the brightest hologram reconstruction, as well as a close-to-forty-five-degree incident angle (fig. 5). This reconstruction geometry was kept unchanged for different holograms and characterization techniques in the following discussions.

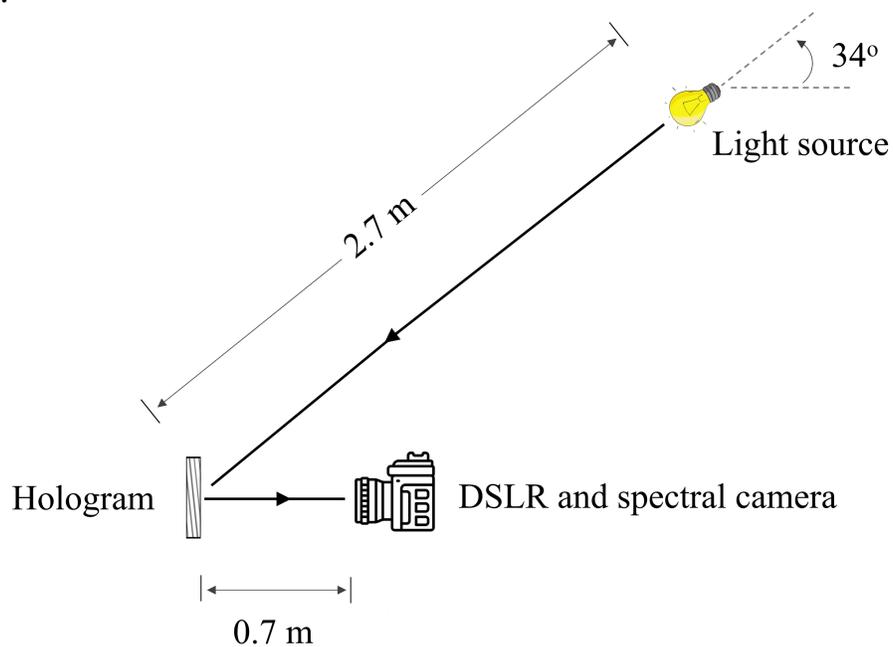

*Figure 5. Diagram of the reconstruction geometry.*

With the reconstruction setup described above, the hologram series displayed the expected dynamic 3D view at varying viewing positions. To document the visual appearance, we produced a light field of the holograms using a DSLR (digital single-lens reflex) camera attached to an automated *x-y* stage (190 cm by 130 cm). The camera was used to capture a thirteen-by-eight array of viewing angles such that, for each of the 104 images, the camera position was staggered from all surrounding positions by fifteen centimeters in both the vertical and horizontal directions (see figs. 6A and 6B for two of the 104 viewing angles). Once all images were synthesized into a single video or GIF file, we obtained a dynamic rendering of the train. Out of practicality, not all possible angles from the holographic scene were captured using this approach. However, enough of the information is preserved from these light-field arrays to be considered as a promising approach toward conservation documentation.

While the light field records the vertical and horizontal parallax of the holograms, we sought to further analyze the scene to infer Benton's creative processes. In particular, we wanted to estimate the optical arrangement used in recording the holograms and how Benton positioned the object relative to the plane of the hologram. To do this, we constructed a 3D model of the holographic scene using the principles of photogrammetry, which has become more common in the field of cultural heritage, ranging in application from large-scale archaeological sites to small objects.[25] Similarly, here we produced a 3D model from the MIT Museum's *Engine no. 9* copy using commercial software (fig. 6C).[26] Because of the speckle inherent to the hologram combined with the attenuation of light from the small camera aperture, which was required to capture a full depth of field, the 3D model exhibits considerable noise. Despite this limitation, the model records the general holographic space and appears in red with the actual plane of the holographic film demarcated by the black edges of the mount and the fiducial markers. Notably, in figure 6C the reconstructed 3D train straddles this plane, suggesting that Benton utilized a technique known as "image plane holography." Rather than using a real train that would reconstruct behind the hologram in these experiments, Benton projected a master hologram (called H1) onto a holographic film (H2) and controlled the position of the projection, so that the reconstructed space would appear in front of the hologram. The technique augments the visual experience of these holograms by projecting the train out toward the viewer, and it was commonly applied by holographers to more easily control brightness and contrast as well as to create sharper reconstruction in reflection holograms.[27]

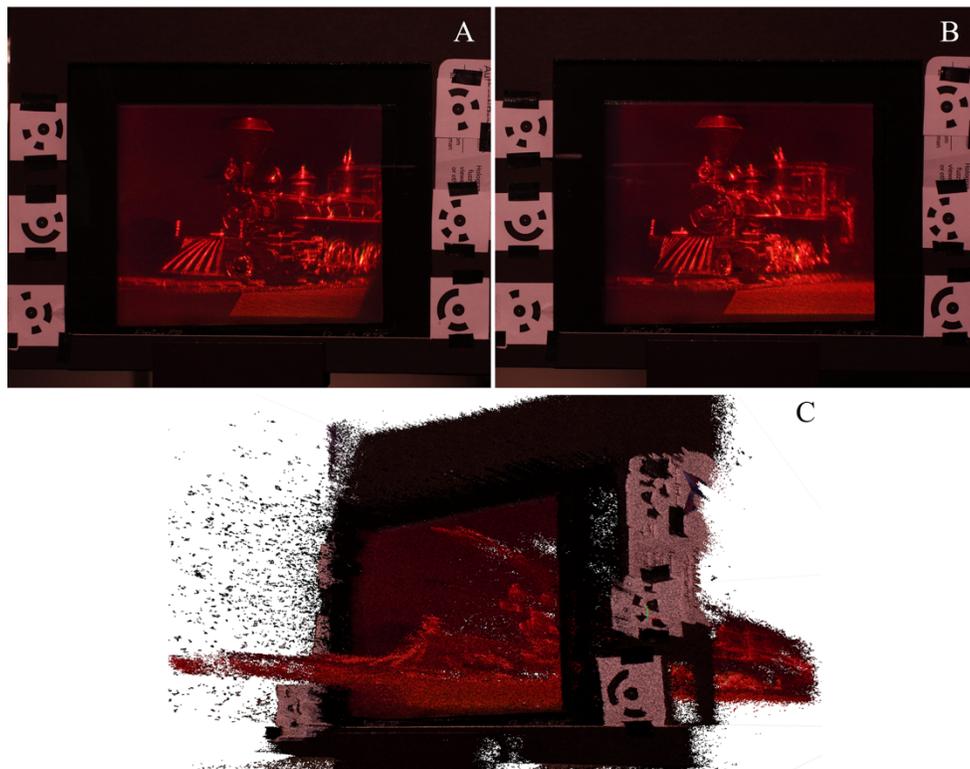

*Figure 6. Registered DSLR images of MIT Engine no. 9 copy at two camera positions (A, B), providing different viewing angles of the train. Six black-and-white fiducial markers along the borders of the hologram assisted with image registration. C, the dense cloud 3D model of MIT Engine no. 9 copy rendered by Agisoft Metashape Professional Edition. The model shows that the plane of the reconstructed wavefront straddles the hologram plane.*

When we viewed the *Engine no. 9* series by eye (fig. 7), the varying colors among prints became obvious and prompted us to quantitatively compare the color of each of the twelve copies by spectral analysis. With a spectral camera (Resonon Pika II), we obtained the reflectance spectra over the entire hologram, then averaged the spectra at all pixels for each copy (fig. 8). Unlike most reflection holograms, which possessed a greenish tint caused by the shrinkage of the emulsion, nine of the twelve *Engine no. 9* copies exhibited different shades of red (peak centered from 617 nm to 647 nm).[28] This maintenance of the red color close to the wavelength of the He-Ne recording laser at 633 nm is likely the result of Benton's use of the IEDT bleach in this series. It is also interesting to note that in each copy, the noise level (the baseline) and brightness (the peak height) varied as well. Among all copies, the MIT copy possessed the highest brightness with lowest noise—possibly why it was selected for donation to the museum as the best quality from the series.

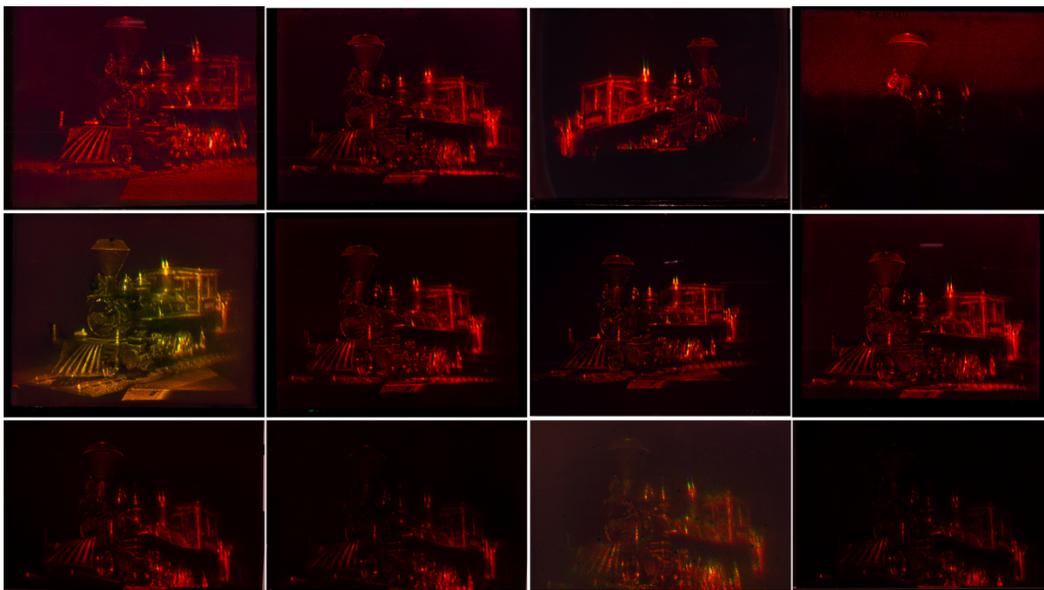

*Figure 7. The Engine no. 9 series under reconstruction. The twelve copies varied in brightness, noise level, and color. Nine of them displayed different shades of red. While the naked eye could hardly detect the differences, a noninvasive quantitative analysis was possible with spectral imaging.*

In addition to the shape and color of the holographic scene, the physical materials of the hologram also matter: the chemical compositions and the microstructures of a hologram that might deteriorate over time. Given these factors, spectral imaging again stood out as a method whereby we could noninvasively estimate some of these physical characteristics. In particular, since a hologram's thickness is correlated to color, changes of thickness due to swelling and shrinkage across the image space could be mapped. As observed on one of the copies, the reflectance spectra from distinct regions (fig. 9) indicated inconsistent thickness. Most significantly, the area of tidelines possessed thicker emulsion or larger fringe spacings compared with the rest of the regions, indicated by the seven-nanometer redshift of the peak position.[29] According to the Bragg equation, a spectral shift of seven nanometers corresponds to a change in the emulsion thickness of approximately one percent or two hundred nanometers, suggesting its ability to detect a ~seventy nanometer change of thickness at the resolution limit. In fact, the tideline was hypothetically caused by water vapor diffusion through the edges of the hologram, during which the emulsion swelled. The spectral analysis therefore indicated the possibility of detecting such thickness change with time and across different regions of the hologram; a pixel-wise analysis would help to map this change quantitatively.

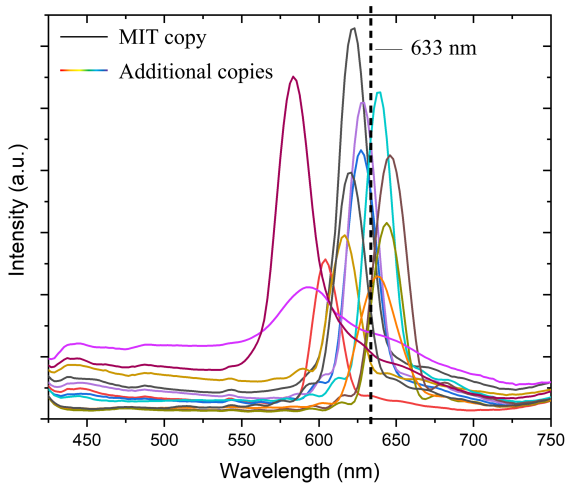

*Figure 8. Average spectra of twelve copies of Engine no. 9.*

**Further Thoughts**

In this brief case study, we proposed the use of light-field captures and spectral imaging to characterize optical properties of reflection holograms. The light-field captures, a potential documentation approach for holograms, faithfully recorded and displayed the vertical and horizontal parallax of Benton's *Engine no. 9* series as well as preserving their depth of field. Combined with geometric analysis using photogrammetry, the bisecting holographic space with the hologram plane suggested that one of the recording techniques used by Benton was image plane holography. Through spectral imaging, the color variation in different *Engine no. 9* copies suggested that Benton experimented with chemical processing procedures. Particularly, spectral analysis was sensitive to the change of thickness at a submicron scale, thus detecting regional swelling or shrinking of the emulsion due to degradation and/or ingress of water vapor. Thus, the two analytical techniques used here offer a glimpse into Benton's efforts in making reflection holograms—experimenting with chemical and optical techniques and pushing the technical limitations—to produce the MIT copy, which is described as "an exceptionally bright and unusually deep example of its kind."[30]

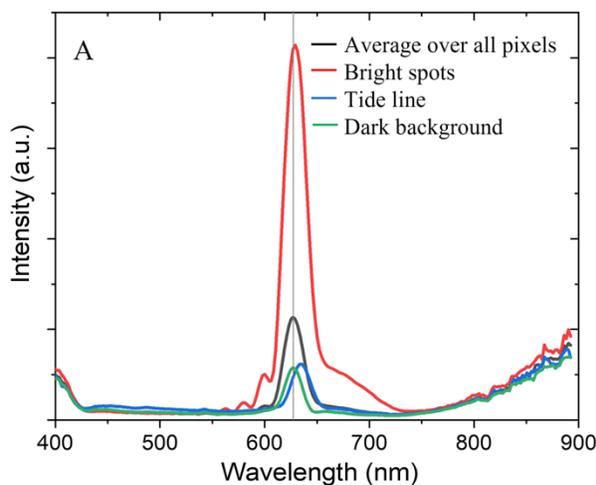
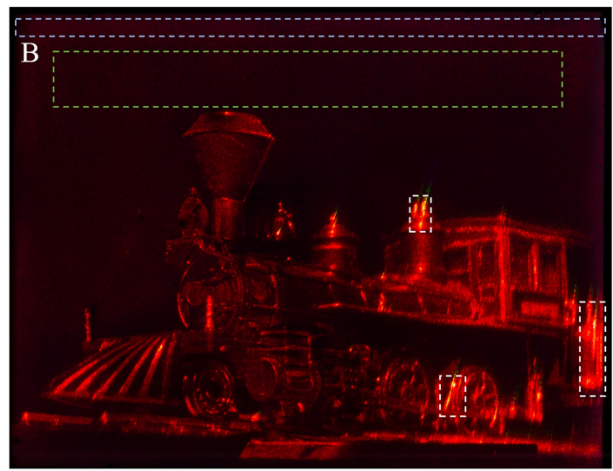

*Figure 9. Spectral analysis of Benton 12. A, the spectra of the hologram averaged in different regions: black—all pixels, red—bright spots, blue—the edges, and green—the dark background area; B, a DSLR capture shows the regions corresponding to the averaged spectra.*

Nevertheless, the current study requires optimization of experimental procedures to improve the quality of the data. In our current research we are optimizing both illumination and the light-field capture process. For instance, implementing a vibrating light source would theoretically reduce laser speckle. Synthesizing a large depth of field by applying focus stacking, combined with optimizing camera positions, would improve resolution. Also, as this type of photography only captures a small portion of the information in a hologram, we are exploring the use of more data-intensive methods, such as laser ptychography, to capture the full wavefront at a resolution of one to two microns. The resulting dataset from these experiments is necessarily large, bordering on hundreds of gigabytes. While such approaches may not be practical to implement or duplicate in many museums' imaging studios, this proof of concept is necessary if full digital holography, with the same resolution as film, is to be realized.

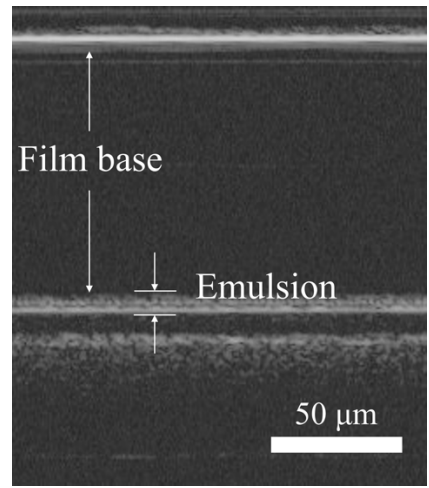

In addition to hyperspectral imaging and light-field captures, analytical techniques such as XRF and OCT could also provide chemical and physical information. For instance, XRF point analysis detected silver in all *Engine no. 9* copies, while bromide and trace elements such as iron were present in some copies. Such elemental information may assist in identifying the type of chemical processing used by the holographer. However, since glass strongly adsorbs X-ray, it blocks the detection of trace elements on the copies layered between glass plates. Therefore, an area without glass substrates or film-based holograms would be more suitable for XRF measurements. OCT, on the other hand, could noninvasively characterize physical layer structures. A preliminary experiment (fig. 10) suggested that a nine-hundred-nanometer OCT system could measure a six-micrometer-thick silver halide emulsion. Since certain brands and types of a holographic film or plate have a defined layered structure, OCT may prove to be valuable for such identification.

*Figure 10. Traverse scan (B-scan) of a holographic film using OCT generated a cross-sectional image. It noninvasively measures the thickness of the emulsion and the film base. The image was obtained using a Thorlabs Ganymede GAN621 900 nm OCT, with the OCT-LK2-BB lens kit.*

**Conclusion**

Holography was a scientific and cultural symbol from the 1960s to the 1980s. In the case study presented here from Benton's *Engine no. 9*, we characterized the materials and optical properties using hyperspectral imaging and light-field capture and show how these data may be used to recover Benton's experimental methods as he attempted to improve the quality of reflection holograms. While improvements in experimental and analytical procedures are needed, we hope this preliminary work may inspire interest, future studies, and expanded discussions on the conservation of holograms.

**Notes**

The authors thank Deborah Douglas, director of collections and curator of science and technology at the MIT Museum; Gloria Martinez, collections manager at the MIT Museum; and Katie Porter, registrar at the MIT Museum, for preparing for lab space and providing access to the hologram collection at the MIT Museum. The authors thank Betsy Connors, holographic artist and previous colleague of Stephen Benton, for consultation on Benton holograms. The research is made possible through the support of the Andrew Mellon Foundation and the National Science Foundation–Partnerships for International Research and Education (NSF–PIRE) grant. This work made use of the MatCI Facility supported by the MRSEC program of the National Science Foundation (DMR-1720139) at the Materials Research Center of Northwestern University.

[1] Lucas, *Star Wars: A New Hope: The Illustrated Screenplay*, 25; Johnston, "Introduction," 2.

[2] "[Cartier] Hand and Jewels"; Slesin, "Diamond Sleight of Hand."

[3] Connors, Benton, and Seamans, "Report from the MIT Museum."

[4] "Huddle."

[5] "Getty Museum Announces Donation."

[6] Wise, "'Transparent Things'" (2013); Wise, "'Transparent Things,'" (2016); Serra and Moreno, "Experiencias en la conservación"; Brown and Jacobson, "Archival Permanence"; Jacobson and Brown, "Archival Properties."

[7] "Untitled, no. 5 of 8."

[8] Nelson, "Phantom Limbs."

[9] Bavor, "Project Starline."

[10] Slesin, "Diamond Sleight of Hand."

[11] Benton and Bove, "In-Line 'Denisyuk' Reflection Holography."

12 Benton and Bove, "Holograms and Perception," 7.

13 Bjelkhagen, *Silver-Halide Recording Materials*.

14 Johnston, "Three-Dimensional Wavefront," 111.

15 Bjelkhagen, "After-Treatment."

16 Bjelkhagen, "Development."

17 Benton and Bove, "In-Line 'Denisyuk' Reflection Holography"; Bjelkhagen, "Bleaching."

18 Spierings, "Chasing the Holoprinter"; "Holography Pioneer."

19 Johnston, "Americanization of Reflection Holography," 200–202.

20 Burns and Jackson, *Holografi*, 61; Benton, "Intra-Emulsion Diffusion-Transfer" (1978).

21 Bjelkhagen, "Development"; Bjelkhagen, "Bleaching."

22 Benton, "Intra-Emulsion Diffusion-Transfer Processing" (1974); Benton, "Intra-Emulsion Diffusion-Transfer" (1978); Benton, Halle, and Bove, "Practical Issues."

23 Bjelkhagen, "Development"; Benton, "Intra-Emulsion Diffusion-Transfer" (1978).

24 Bjelkhagen and Brotherton-Ratcliffe, "Illumination of Colour Holograms."

25 Aicardi et al., "Recent Trends."

26 "Engine #9"; "AgiSoft PhotoScan Professional."

27 Brandt, "Image Plane Holography."

28 Burns and Jackson, *Holografi*, 61.

29 Trochtchanovitch, "Method of Characterization"; Benton and Bove, "In-Line 'Denisyuk' Reflection Holography."

30 Burns and Jackson, *Holografi*, 61.